\documentclass[11pt,twoside]{article}

\usepackage{asp2006}
\usepackage{epsf}
\usepackage{psfig}
\usepackage{lscape}

\begin{document}
\title{The Mass of Virialized Dark Matter Haloes}
\author{Antonio J. Cuesta and Francisco Prada}
\affil{Instituto de Astrof\'{\i}sica de Andaluc\'{\i}a (CSIC), Spain}

\bibliographystyle{natbib}

\begin{abstract}
Virial mass is used as an estimator for the mass of a dark matter halo. However, the commonly used constant overdensity criterion does not reflect the dynamical structure of haloes. Here we analyze dark matter cosmological simulations in order to obtain properties of haloes of different masses focusing on the size of the region with zero mean radial velocity. Dark matter inside this region is stationary, and thus the mass of this region is a much better approximation for the virial mass.
\end{abstract}

Many results of simulations implicitly use some definition of what is the size and mass of a collapsed dark matter halo. The problem is that there is no well-defined boundary: density field is smooth around the halo. The common prescription for this boundary (and hence the mass belonging to the halo) is defined through the spherical collapse model (see references \citealt{Gun72}, \citealt{Gun77}). Thus, it is very common to measure the mass of haloes in cosmological simulations taking the particles inside a sphere of fixed spherical overdensity, whose radius is often misnamed the \emph{virial radius}.\\

\begin{figure}[!ht]
\plottwo{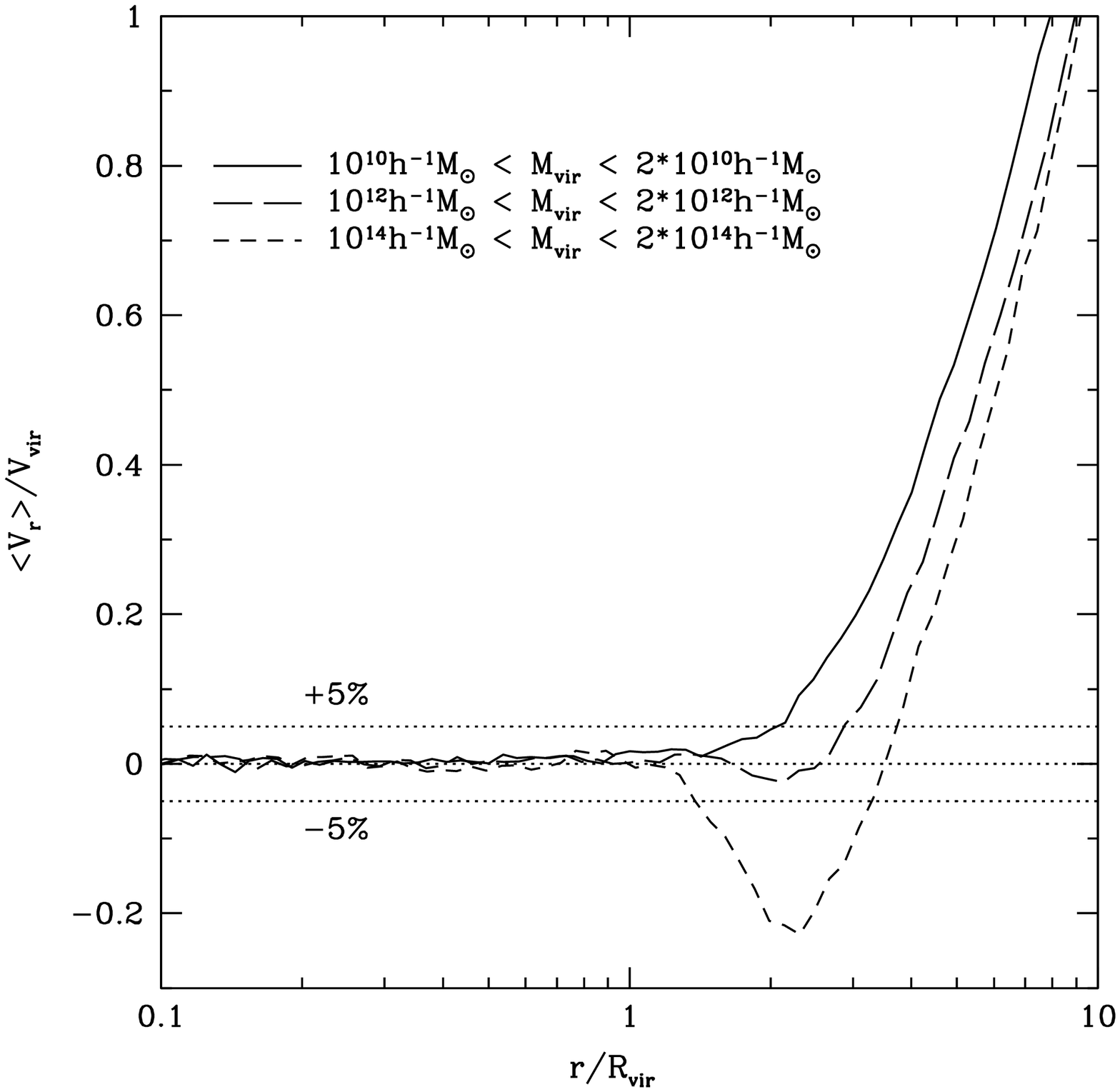}{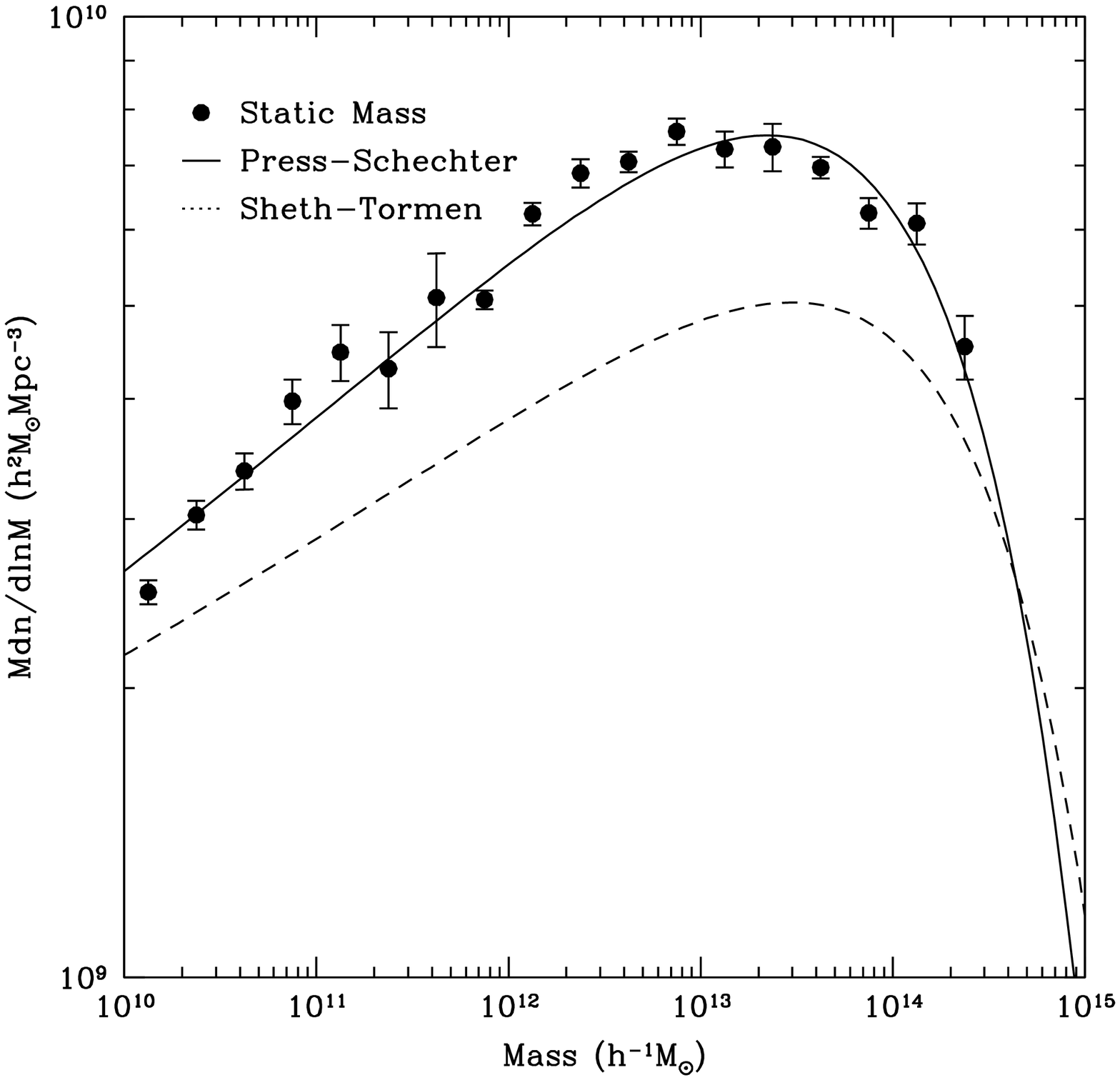}
\caption{Left panel: Mean radial velocity for three different mass bins. The profiles were obtained by averaging over hundreds of haloes on each mass bin. In dotted line is shown the selected threshold delimiting the static region (5 \% of the virial velocity). Cluster-size haloes display a region with strong infall (dashed line). On the contrary, low-mass haloes (solid line) and galactic haloes (long-dashed line) do not show infall at all. Right panel: The static mass function (solid circles). Error bars show the Poissonian error only. Solid and dashed lines show the P-S and S-T mass function, respectively.}
\label{fig}
\end{figure}

The structure of haloes in phase-space, already discussed in \cite{Bus05}, is as follows: In the inner parts the average radial velocities are zero. Around the virial radius there are some signs of infall or outflow, which strongly correlate with the mass scale \citep{Pra06}. At large distances, the Hubble flow is the most remarkable feature. This structure shows a substantial difference in dark matter haloes associated with different masses (see left panel in Figure~\ref{fig}). Although the transition from the inner region to the Hubble flow shows monotonically increasing radial velocity in the low-mass and the galactic-size halo, the situation is different for clusters where there is a significant tail in the velocity distribution with large negative velocities. This infall is altering the equilibrium of the particle distribution in the halo, and hence that region is no longer virialized.\\

This suggests a natural choice for the boundary of a dark matter halo, which is defined by the innermost radius in which mean radial velocity is equal to zero. In practice, we use a small threshold in order to reduce effects of the statistical noise. The deviation of the spherically averaged radial velocity profile from zero is a signature of non-virialization. For this reason the zero-velocity radius is defining the mass which is associated to a halo. This radius is referred to as the static radius, and the mass inside a sphere of this radius, as the static mass of the halo.\\

In spite of the fact that the mass function of \cite{She99} provides a very good fit for the virial mass, we find that this is no longer a good description for the static mass function (see right panel in Figure~\ref{fig}). Surprisingly, the mass function of \cite{Pre74} approximates really well our data in the mass range under study.

Please refer to arXiv:0710.5520 (Cuesta et al.) for further details.

\acknowledgements
A.J.C., F.P. thank the Spanish MEC under grant PNAYA 2005-07789 for their support. A.J.C. acknowledges the financial support of the MEC through Spanish grant FPU AP2005-1826. Computer simulations were done at the LRZ Munich, NIC Julich, and NASA Ames.

\bibliography{mycites}

\begin{thebibliography}{}

\bibitem[{Busha} {et~al.}(2005){Busha}, {Evrard}, {Adams}, and
  {Wechsler}]{Bus05}
{Busha}, M.~T., {Evrard}, A.~E., {Adams}, F.~C., and {Wechsler}, R.~H. (2005).
\newblock {The ultimate halo mass in a {$\Lambda$}CDM universe}.
\newblock {\em \mnras\/}, {\bf 363}, L11--L15.

\bibitem[{Gunn}(1977){Gunn}]{Gun77}
{Gunn}, J.~E. (1977).
\newblock {Massive galactic halos. I - Formation and evolution}.
\newblock {\em \apj\/}, {\bf 218}, 592--598.

\bibitem[{Gunn} and {Gott}(1972){Gunn} and {Gott}]{Gun72}
{Gunn}, J.~E. and {Gott}, J.~R.~I. (1972).
\newblock {On the Infall of Matter Into Clusters of Galaxies and Some Effects
  on Their Evolution}.
\newblock {\em \apj\/}, {\bf 176}, 1.

\bibitem[{Prada} {et~al.}(2006){Prada}, {Klypin}, {Simonneau},
  {Betancort-Rijo}, {Patiri}, {Gottl{\"o}ber}, and {Sanchez-Conde}]{Pra06}
{Prada}, F., {Klypin}, A.~A., {Simonneau}, E., {Betancort-Rijo}, J., {Patiri},
  S., {Gottl{\"o}ber}, S., and {Sanchez-Conde}, M.~A. (2006).
\newblock {How Far Do They Go? The Outer Structure of Galactic Dark Matter
  Halos}.
\newblock {\em \apj\/}, {\bf 645}, 1001--1011.

\bibitem[{Press} and {Schechter}(1974){Press} and {Schechter}]{Pre74}
{Press}, W.~H. and {Schechter}, P. (1974).
\newblock {Formation of Galaxies and Clusters of Galaxies by Self-Similar
  Gravitational Condensation}.
\newblock {\em \apj\/}, {\bf 187}, 425--438.

\bibitem[{Sheth} and {Tormen}(1999){Sheth} and {Tormen}]{She99}
{Sheth}, R.~K. and {Tormen}, G. (1999).
\newblock {Large-scale bias and the peak background split}.
\newblock {\em \mnras\/}, {\bf 308}, 119--126.

\end{thebibliography}

\end{document}